\documentclass[preprint,pra,showpacs,showkeys]{revtex4}
\usepackage{graphicx,amsmath,amsfonts,amssymb}
\usepackage{times}

\begin{document}

\title{Entanglement control in one-dimensional
$s=\frac{1}{2}$ random XY spin chain
\footnote{Supported by the Key
Higher Education Programme of Hubei Province under Grant No
Z20052201, the Natural Science Foundation of Hubei Province, China
under Grant No 2006ABA055, and the Postgraduate Programme of Hubei
Normal University under Grant No 2007D20.}}

\author{C. J. Shan\footnote{ E-mail: scj1122@163.com}}
\author{W. W. Cheng}
\author{T. K. Liu\footnote{Corresponding author. E-mail:
tkliuhs@163.com}}
\author{Y. X. Huang}
\author{H. Li}
\affiliation{College of Physics and Electronic Science, Hubei Normal
University, Huangshi 435002, China}
\date{\today}

\begin{abstract}

The entanglement in  one-dimensional  random XY spin systems where
the impurities of exchange couplings and the external magnetic
fields are considered as random variables is investigated by solving
the different spin-spin correlation functions and the average
magnetization per spin. The entanglement dynamics near particular
locations of the system is also studied when the exchange couplings
(or the external magnetic fields) satisfy three different
distributions(the Gaussian distribution, double-Gaussian
distribution, and bimodal distribution). We find that the
entanglement can be controlled by varying the strength of external
magnetic field and the different distributions of impurities.
Moreover, the entanglement of some nearest-neighboring qubits can be
increased for certain parameter values of the three different
distributions.
\end{abstract}

\pacs{03.65.Ud, 03.67.Mn, 75.10.Pq}

\keywords{the entanglement, spin chain, spin-spin correlation
function, Gaussian distribution}

\maketitle

\section{Introduction}
Entanglement not only has the interesting properties of quantum
mechanics but also is very important in the quantum information
processing (QIP), such as quantum teleportation$^{[1]}$, dense
coding$^{[2]}$, quantum secret sharing$^{[3]}$, quantum
computation$^{[4]}$ and some cryptographic protocols$^{[5]}$. In
order to realize the quantum information process, a great effort has
been made to study and characterize the entanglement in cavity
QED$^{[6-7]}$ and solid state systems. A typical example is the spin
chains that can describe interaction of qubits not only in solid
physical systems but also in many other systems such as quantum
dots$^{[8]}$, electronic spins$^{[9]}$, and optical
lattices$^{[10]}$. Therefore, there have been conducted numerous
studies on Ising model$^{[11]}$ and all kinds of
Heisenberg XY XXZ XYZ models$^{[12-15]}$.\\
\indent Impurities often exist in solid systems and play an
important part in condensed matter physics. As a candidate of QIP, a
solid system with impurity is also one of our important study
objects. In the previous researches, the effect of impurity  on the
quantum entanglement has been studied in a three-spin system
$^{[16-17]}$ and a large spin systems under zero
temperature$^{[18]}$.
However, in these studies, only single impurity has been studied.\\
\indent Recently, Huang et al$^{[19-20]}$ have demonstrated that for
a class of one-dimensional magnetic systems entanglement can be
controlled and tuned by varying the anisotropy parameter in the XY
Hamiltonian and by introducing impurities into the systems. However,
in Ref.(19), only the impurity and the external magnetic fields in a
Gaussian form are considered and the value of the width of the
distribution is fixed. In Ref.(20), the strength of the impurity is
located at two sites.  For the pure case, Osterloh et al$^{[21]}$
examined the entanglement between two spins at position i and j in
the spin chains. Owing to its importance, in this paper we study the
entanglement dynamics near particular locations of one-dimensional
$s=\frac{1}{2}$ random XY spin system when the exchange couplings
(or external magnetic fields) satisfy three different
distributions(the Gaussian distribution,  double-Gaussian
distribution, and bimodal distribution), to our knowledge, which
have not been reported yet. The present study
 in simple examples can help us to understand
the behaviour of the entanglement in one-dimensional  random XY spin
systems for the different distributions. More importantly, we will
demonstrate that one can control or manipulate the entanglement in
spin system with the help of the exchange couplings and the external
magnetic fields.
\section{Solution of the XY model}
We consider a physical Heisenberg XY model of N spin-$\frac{1}{2}$
particles interacting with their nearest neighbours. In the presence
of impurities, the one-dimensional Hamiltonian is given by $^{[19]}$
\begin{eqnarray}
H=&-&\frac{1+\gamma
}{2}\sum_{i=1}^{N}J_{i,i+1}\sigma _{i}^{x}\sigma
_{i+1}^{x}-\frac{1-\gamma
}{2}\nonumber\\&\times&\sum_{i=1}^{N}J_{i,i+1}\sigma _{i}^{y}\sigma
_{i+1}^{y}-\sum_{i=1}^{N}h_{i}\sigma _{i}^{z}
\end{eqnarray}
where $J_{i,i+1}$ is the exchange interaction between sites i and
i+1, $h_{i}$ is the strength of the external magnetic field on site
i,  $\sigma^{x,y,z}$ are the Pauli matrices,  $\gamma$ is the degree
of anisotropy and N is the number of sites. The periodic boundary
conditions satisfy $\sigma _{N+1}^{x}=\sigma _{1}^{x}, \sigma
_{N+1}^{y}=\sigma _{1}^{y}, \sigma _{N+1}^{z}=\sigma
_{1}^{z}$.\\
 Now we define the raising and lowing operators $a_{i}^{+}$,$a_{i}^{-}$ and introduce Fermi
 operators$^{[22]}$
 $c_{j}^{+}$ and $c_{j}$, and they are expressed as follows:\\
 \begin{eqnarray}a_{i}^{+}=\frac{1}{2}(\sigma _{i}^{x}+i\sigma
_{i}^{y})=c_{i}^{+}\exp (i\pi \sum_{j=1}^{i-1}c_{j}^{+}c_{j}), \\
a_{i}^{-}=\frac{1}{2}(\sigma _{i}^{x}-i\sigma _{i}^{y})=\exp (-i\pi
\sum_{j=1}^{i-1}c_{j}^{+}c_{j})c_{i}.
\end{eqnarray}
so that, the Hamiltonian has the following form
\begin{eqnarray}
H=&-&\sum_{i=1}^{N}J_{i,i+1}[(c_{i}^{+}c_{i+1}+\gamma
c_{i}^{+}c_{i+1}^{+})+h.c]\nonumber\\&-&2\sum_{i=1}^{N}h_{i}(c_{i}^{+}c_{i}-\frac{
1}{2})
\end{eqnarray}
In the present paper, the exchange interaction has the form
$J_{i,i+1}=J(1+\alpha_{i,i+1})$, where $\alpha$ introduces the
impurity in the double-Gaussian form with peaks at $\frac{N+1}{2}$
with strength $\zeta _{1}$ and  at $\frac{N-1}{2}$ with strength
$\zeta _{2}$,
\begin{eqnarray}
\alpha _{i,i+1}&=&P\times \zeta _{1}\exp \{-\epsilon (i-\frac{N+1}{2}%
)\}\nonumber\\&+&(1-P)\times \zeta _{2}\exp \{-\epsilon
(i-\frac{N-1}{2})\}
\end{eqnarray}
 The external magnetic field
takes the form $h_{i}=h(1+\beta _{i})$, where
\begin{eqnarray}
\beta _{i}&=&P\times \xi _{1}\exp \{-\epsilon
(i-\frac{N+1}{2})\}\nonumber\\&+&(1-P)\times \xi _{2}\exp
\{-\epsilon (i-\frac{N-1}{2})\}
\end{eqnarray}
When $\alpha=\beta=0$, the above reduces to a  pure case; when
$P=1$, the above reduces to the case in Ref.(19).  By introducing
the dimensionless parameter $\lambda=J/(2h)$, the symmetrical matrix
A and the antisymmetrical B, the Hamiltonian becomes
\begin{eqnarray}
H=\sum_{i,j=1}^{N}[c_{i}^{+}A_{i,j}c_{j}+\frac{1}{2}%
(c_{i}^{+}B_{i,j}c_{j}^{+}+h.c)]
\end{eqnarray}
The above Hamiltonian can be diagonalized by making linear
transformation of the fermionic operators $\eta
_{k}=\sum_{i=1}^{N}g_{ki}c_{i}+h_{ki}c_{i}^{+},\eta
_{k}^{+}=\sum_{i=1}^{N}g_{ki}c_{i}^{+}+h_{ki}c_{i},$  and then the
Hamiltonian becomes
\begin{eqnarray}
H=\sum_{k=1}^{N}\Lambda _{k}\eta _{k}^{+}\eta _{k}+const,
\end{eqnarray}
and two coupled  matrix equations satisfy $\phi _{k}(A-B)=\Lambda
_{k}\psi _{k},\psi _{k}(A+B)=\Lambda _{k}\phi _{k},$ where the
components of the two column vectors $\phi _{ki},\psi _{ki}$ are
given by $\phi _{ki}=g_{ki}+h_{ki},\psi _{ki}=g_{ki}-h_{ki}.$
Finally, the ground state of the system $\left\vert \psi
_{0}\right\rangle $ can be written as $\eta _{k}\left\vert \psi
_{0}\right\rangle =0$.
\section{Spin-spin correlation functions}
\indent Before we dicuss the entanglement, we should have a brief
review of spin-spin correlation functions. The spin-spin correlation
functions for ground state and the average magnetization per spin
are respectively defined
as $^{[22]}$\\

$S_{lm}^{x}=\frac{1}{4}\left\langle \psi _{0}\left\vert \sigma
_{l}^{x}\sigma _{m}^{x}\right\vert \psi _{0}\right\rangle,
S_{lm}^{y}=\frac{1}{4}\left\langle \psi _{0}\left\vert \sigma
_{l}^{y}\sigma _{m}^{y}\right\vert \psi _{0}\right\rangle ,$\\

$S_{lm}^{z}=\frac{1}{4}\left\langle \psi _{0}\left\vert \sigma
_{l}^{z}\sigma _{m}^{z}\right\vert \psi _{0}\right\rangle ,
M_{i}^{z}=\frac{1}{2}\left\langle \psi _{0}\left\vert \sigma
_{i}^{z}\right\vert \psi _{0}\right\rangle .$\\

These correlation functions are given as expectation values of
products of fermion operators. Using Wicks theorem$^{[23]}$, these
expressions can be rewritten as \\

$S_{lm}^{x}=\frac{1}{4}\left(
\begin{array}{cccc}
G_{l,l+1} & G_{l,l+2} & \cdots  & G_{l,m} \\
\vdots  & \vdots  & \ddots  & \vdots  \\
G_{m-1,l+1} & G_{m-1,l+1} & \cdots  & G_{m-1,m}%
\end{array}%
\right)$,\\

$S_{lm}^{y}=\frac{1}{4}\left(
\begin{array}{cccc}
G_{l+1,l} & G_{l+1,l+1} & \cdots  & G_{l+1,m-1} \\
\vdots  & \vdots  & \ddots  & \vdots  \\
G_{m,l} & G_{m,l+1} & \cdots  & G_{m,m-1}%
\end{array}%
\right) $\\

$S_{lm}^{z}=\frac{1}{4}(G_{l,l}G_{m,m}-G_{m,l}G_{l,m}),M_{i}^{z}=\frac{1}{2}%
G_{i,i}$ \\

where $G_{i,j}=-\sum_{k}^{N}\psi _{ki}\phi _{kj}$
\section{Entanglement of nearest-neighbouring qubits }

In this part, we give the expression of the concurrence that
quantifies the amount of the entanglement between two qubits.
 For a system described by the density matrix$\rho$, the concurrence C
is$^{[24]}$
\begin{eqnarray}
C(\rho )=\max (0,\lambda _{1}-\lambda _{2}-\lambda _{3}-\lambda
_{4})
\end{eqnarray}
Here $\lambda _{1}$, $\lambda _{2}$, $\lambda _{3}$, $\lambda _{4}$
are the eigenvalues (of them $\lambda _{1}$ is the largest) of the
spin-flipped density operator R, which is defined by
 $R=\sqrt{\sqrt{\rho }\tilde{\rho} \sqrt{\rho }}$
, where $\tilde{\rho} =(\sigma _{y}\otimes \sigma _{y})\rho ^{\ast
}(\sigma _{y}\otimes \sigma _{y})$, $\tilde{\rho}$ denoting the
complex conjugate of $\rho$ with $\sigma _{y}$ being the usual Pauli
matrix. The values of concurrence C ranges from zero to one; when
$C=0$, the two qubits are in  an unentangled state, when $C=1$, the
two qubits are in an maximally entangled state.\\
\indent Using the operator expansion for the density matrix and the
symmetries of the Hamiltonian$^{[25]}$, in the basis states
$\{\left\vert \uparrow \uparrow \right\rangle ,\left\vert \uparrow
\downarrow \right\rangle ,\left\vert \downarrow \uparrow
\right\rangle ,\left\vert \downarrow \downarrow \right\rangle \}$, $\rho$ has the general form \\

$\rho=\left(
\begin{array}{cccc}
\rho _{1,1} & 0 & 0 & \rho _{1,4} \\
0 & \rho _{2,2} & \rho _{2,3} & 0 \\
0 & \rho _{3,2} & \rho _{3,3} & 0 \\
\rho _{4,1} & 0 & 0 & \rho _{4,4}%
\end{array}%
\right) ,$\\
with \\
$\lambda _{a}=\sqrt{\rho _{1,1}\rho _{4,4}}+\left\vert \rho
_{1,4}\right\vert ,\lambda _{b}=\sqrt{\rho _{2,2}\rho
_{3,3}}+\left\vert \rho _{2,3}\right\vert ,$\\
$\lambda _{c}=\sqrt{\rho _{1,1}\rho _{4,4}}-\left\vert \rho
_{1,4}\right\vert ,\lambda _{d}=\sqrt{\rho _{2,2}\rho
_{3,3}}-\left\vert \rho _{2,3}\right\vert ,$

We can express all the matrix elements in the density matrix in
terms of different spin-spin correlation functions:\\

$\rho _{1,1}=\frac{1}{2}M_{l}^{z}+\frac{1}{2}M_{m}^{z}+S_{lm}^{z}+\frac{1}{4}%
,$

$\rho _{2,2}=\frac{1}{2}M_{l}^{z}-\frac{1}{2}M_{m}^{z}-S_{lm}^{z}+\frac{1}{4}%
,$

$\rho _{3,3}=\frac{1}{2}M_{m}^{z}-\frac{1}{2}M_{l}^{z}-S_{lm}^{z}+\frac{1}{4}%
,$

$\rho _{4,4}=-\frac{1}{2}M_{l}^{z}-\frac{1}{2}M_{m}^{z}+S_{lm}^{z}+\frac{1}{4%
},$

$\rho _{2,3}=S_{lm}^{x}+S_{lm}^{y},$

$\rho _{1,4}=S_{lm}^{x}-S_{lm}^{y},$ \\
\section{Results and discussions }

In this paper, we focus our discussions on the transverse Ising
model with $\gamma=1$. Our goal is to examine the dynamics of
entanglement in the  varying of the exchange couplings and
 the external magnetic fields.  First, we examine the change of
the entanglement for the nearest neighbouring concurrence
C(i,i+1)for different values of the impurity as the parameter
$\lambda$ varies. We consider two kinds of nearest neighbouring
concurrences near particular locations of the system. Figure 1
depicts the nearest neighbouring concurrence C(49,50) as a function
of the reduced coupling constant $\lambda$ at different values of
the impurity $\zeta$ for different distributions with the system
size N =101 and the anisotropy parameter $\gamma=1$. Figure 1(a)
shows the change of concurrence C(49,50)  as a function of different
values $\lambda$ with $p=1$, i.e the Gaussian distribution. We can
see that the concurrence  increases  and arrives at a maximum close
to the critical point $\lambda_{c}$, and it is close to zero above
$\lambda_{c}$. As $\zeta$ increases the concurrence tends to
increase faster and the $\lambda_{m}$, where concurrence approaches
a maximum, shift to left very rapidly. This is consistent with the
result in Ref. [19](Fig.1). In Fig.1(b), 1(c), and 1(d), we give the
curves for the concurrence against the width of the double-Gaussian
distribution. The two Gaussian distributions have equal probability
with  $p=0.5$, and the central positions are at $\frac{N+1}{2}$ and
$\frac{N-1}{2}$. Here one of the double-Gaussian distribution is
fixed with $\zeta_{1}=0.5$. We first investigate the situation when
the width of the double-Gaussian distribution $\epsilon$ is 0.1, A
similar behaviour can be seen in Fig.1(b), only the changed width
 becomes narrow. As $\epsilon$ increases, the concurrence
increases slowly and the peak value decreases, which is shown in
Fig.1(c). As is well known, bimodal distribution is a particular
case of the double-Gaussian distribution, that is to say, the
double-Gaussian distribution is converted into bimodal distribution
as $\epsilon$ increases. In Fig.1(d), $\epsilon=10$. The numerical
calculations show that concurrence decreases with the increase of
$\zeta_{2}$, which  indicates that the behaviours are very different
from the former cases.\\
\indent In Fig.2, we show the results of the nearest neighbouring
concurrence between the sites 49 and 50, as a function of the
parameter $\lambda$ for different strengths of the external magnetic
field $\xi$. The effect of the external magnetic field $\xi$ in the
Gaussian distribution is also shown in Fig.2(a). However, different
from the effect of  the exchange couplings, the concurrence
increases slowly and tends to move to infinity by increasing the
value of the parameter $\xi$. This is also consistent with the
result in Ref. [19](Fig.1).  A similar behaviour can be seen in
Fig.2(b) for the double-Gaussian distribution, however, with $\xi=1$
and $\xi=10$, it is interesting to find that the entanglement peak
between the nearest neighbours increases to a  value larger than
that in Fig.2(a). As $\xi$ increases, the concurrence increases
rapidly below $\lambda_{c}$, while the concurrence increases slowly
above $\lambda_{c}$. A comparison between the dash curve and the
dash dotted curve in Fig.2(d) shows that the concurrence increases
rapidly and tends to move to infinity by increasing the value of the
parameter $\xi_{2}$, which is different from the results obtained
from the Gaussian distribution and double-Gaussian distribution.
 That is to say, the strong $\xi_{2}$ is helpful to keep the better entanglement for the
bimodal distribution. \\
\indent Up to now we have examined the nearest neighbouring
concurrence C(49,50) with different Gaussian distributions for
purities and strengths of magnetic field. It is interesting to study
the effect of the different Gaussian distributions on the
concurrence for the rest of the sites in the chain. For the Ising
model, a similar analysis can be carried out for the  nearest
neighbouring concurrence C(50,51),the concurrence is located at the
centre of the double-Gaussian distribution. This is demonstrated in
Figs.3 and 4 by the evolutions of the concurrence.
 Figure 3 corresponds to the case in which the exchange couplings are varying,
and the peak of the maximal entanglement becomes larger than that in
Fig.1.  It is the different distributions that lead to considerable
different evolutions of the entanglement, hence the entanglement is
rather sensitive to any small change in the exchange interaction for
the bimodal distribution. As  shown in Eq.(5), for the bimodal
distribution, the strengths of impurity are mostly located at two
sites($\alpha_{49,50}, \alpha_{50,51}$). The nearest neighbouring
concurrence increases with the increasing of $\zeta_{2}$, so that by
adjusting $\zeta_{2}$ one can obtain a strong entanglement. The
results that we have obtained here are also consistent with those in
Ref. [20](Fig.4). Figure 4 corresponds to the case in which the
external magnetic field is varying, the entanglement between nearest
neighbours tends to be reduced in the presence of the external
magnetic field for the double-Gaussian distribution and the bimodal
distribution, while the entanglement between 49 and 50 increases as
shown in Fig.2. The numerical calculations also show that  as the
parameter $\lambda$ increases from 0 to 4, similar behaviours to
those in Figs.2(c) and 2(d) are shown in
Figs.4(c) and 4(d).\\
\indent  From the above analysis, it is clear that the three
different distributions(the Gaussian distribution, double-Gaussian
distribution, and bimodal distribution) have a notable influence on
the nearest neighbouring concurrence. As for the case
$\gamma\neq1$(XY model) or the next nearest neighbouring
concurrence, we will present further reports in the future.
\section{conclusion }
\indent The entanglements  near particular locations in a
one-dimensional $s=\frac{1}{2}$ random XY spin system have been
investigated. Through analyzing  the exchange couplings (or external
magnetic fields) of three different distributions(the Gaussian
distribution, double-Gaussian distribution, and bimodal
distribution), we have shown that the entanglement can be controlled
and enhanced by varying the strengths of the magnetic field and the
impurity distribution in the system.  The nearest neighbouring
concurrence exhibits some interesting phenomena. For a certain
distribution, concurrence C(49,50) decreases with the increase of
$\zeta$, while concurrence C(50,51) increases. Different behaviours
in the varying of the external magnetic field can occur close to and
above the critical point. The different distributions play an
important role in enhancing the entanglement.

\newpage
\begin{figure}
\begin{center}
\includegraphics[width=1.0\textwidth]{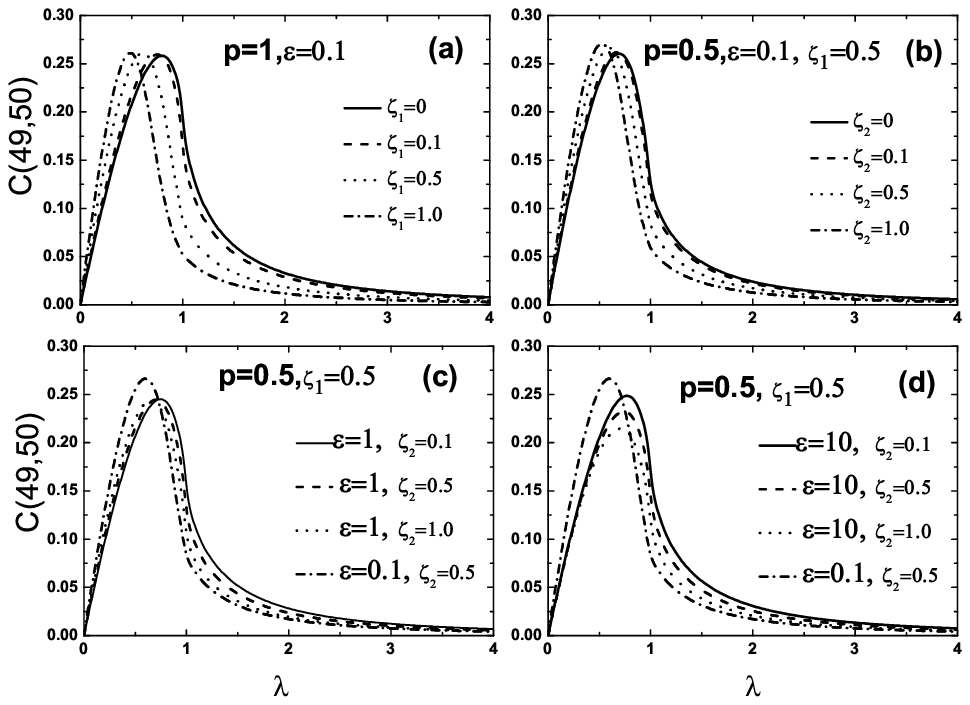}\\
\caption{The nearest neighbouring concurrence C(49,50) as a function
of the reduced coupling constant $\lambda$ at different values of
impurity $\zeta$ for different distributions, with the system size N
=101 and the anisotropy parameter $\gamma=1$.}\label{Fig.1.EPS}
\end{center}
\end{figure}

\begin{figure}
\begin{center}
\includegraphics[width=1.0\textwidth]{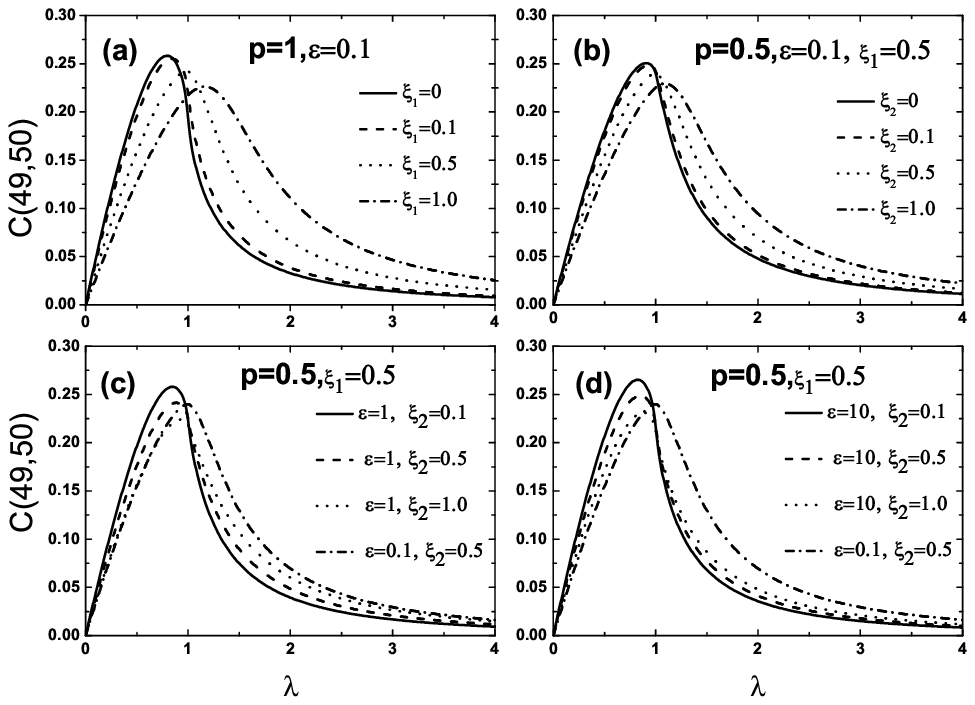}\\
\caption{The nearest neighbouring concurrence C(49,50) as a function
of the reduced coupling constant $\lambda$ different strengths of
the external magnetic field $\xi$, with the system size N =101 and
the anisotropy parameter $\gamma=1$.}\label{Fig.2.EPS}
\end{center}
\end{figure}

\begin{figure}
\begin{center}
\includegraphics[width=1.0\textwidth]{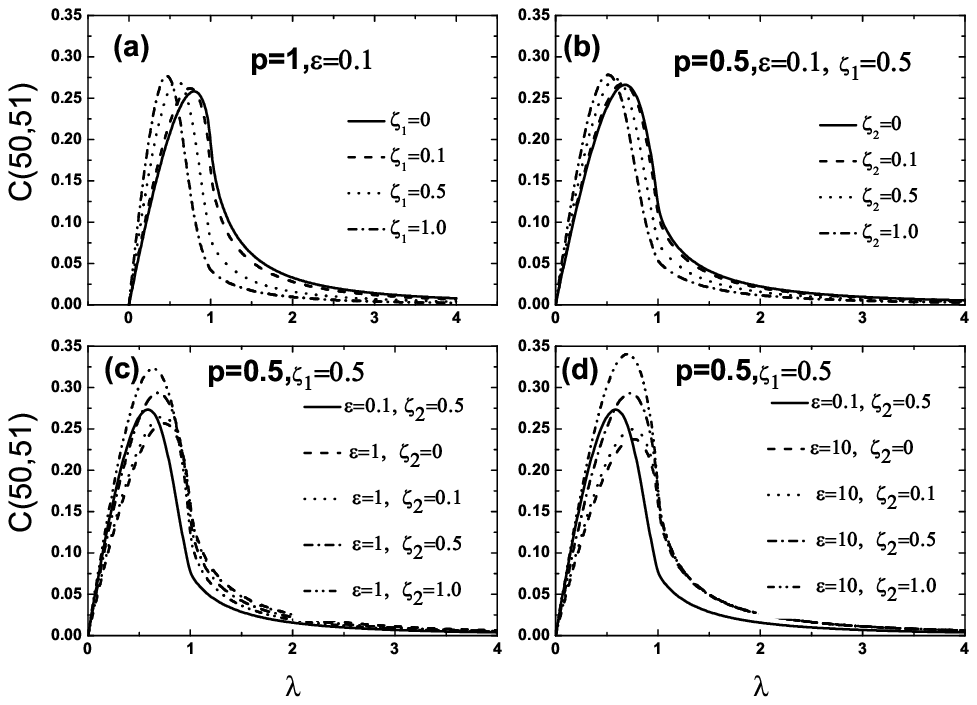}\\
\caption{The nearest neighbouring concurrence C(50,51) as a function
of the reduced coupling constant $\lambda$ at different values of
impurity $\zeta$ for different distributions, with the system size N
=101 and the anisotropy parameter $\gamma=1$.}\label{Fig.3.EPS}
\end{center}
\end{figure}

\begin{figure}
\begin{center}
\includegraphics[width=1.0\textwidth]{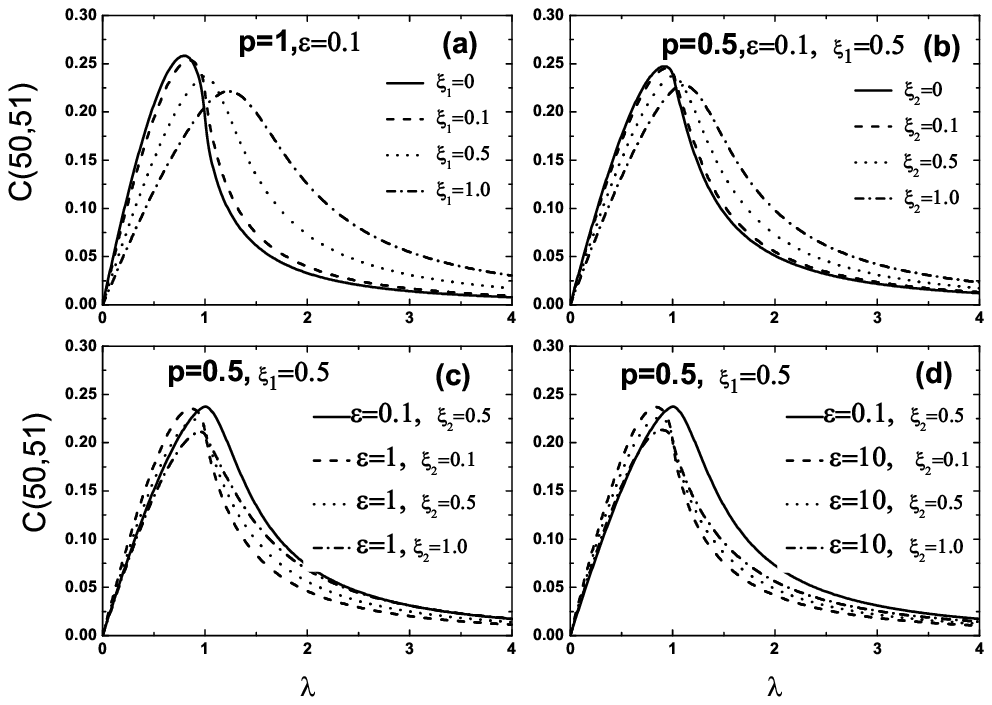}\\
\caption{The nearest neighbouring concurrence C(50,51) as a function
of the reduced coupling constant $\lambda$ different strengths of
the external magnetic field $\xi$, with the system size N =101 and
the anisotropy parameter $\gamma=1$.}\label{Fig.4.EPS}
\end{center}
\end{figure}


\begin{thebibliography}{100}
\bibitem{1} Bennett C H, Brassard G, Cr¨¦peau C, Jozsa R, Peres A, and  Wootters W K, 1993 Phys. Rev. Lett. 70  1895.\\
            Shan C J, Man Z X, Xia Y J, Liu T K. 2007 Int. J. Quantum
              Information, 5 359.
\bibitem{2} Cheng W W, Huang Y X, Liu T K and Li H, 2007 Chin. Phys. 16
38.
\bibitem{3} Shan C J, Man Z X, Xia Y J, Liu T K. 2007 Int. J. Quantum
              Information, 5 335.
\bibitem{4} Grover L 1998 Phys. Rev. Lett. 80 4329.
\bibitem{5} Ma J, Zhang G Y, Rong Y W and Tan L Y 2006 Acta. Phys. Sin. 55 24.\\
            Man Z X, Xia Y J, 2007 Chin. Phys. 16 1197.
\bibitem{6} Shan C J, Xia Y J 2006 Acta. Phys. Sin. 55 1585.
\bibitem{7} Liu T K, 2006 Chin. Phys. 15 0542.\\
            Guo D J, Shan C J, Xia Y J 2007 Acta. Phys. Sin. 56 2139.
\bibitem{8} Loss D, Divincenzo D P,1998 Phys. Rev. A  57 120.
\bibitem{9} Kane B E, 1998 Nature(London) 393 133.
\bibitem{10} Sorensen A, Molmer K, 1999 Phys. Rev. Lett. 82 4556.
\bibitem{11} Wu Y, Machta J, 2005  Phys. Rev. Lett. 95 137208.
\bibitem{12} Wang X G 2001 Phys. Rev. A 64 012313.
\bibitem{13} Zhang G F and Li S S, 2005 Phys. Rev. A  72  034302.
\bibitem{14} Asoudeh M and  Karimipour V, 2005  Phys. Rev. A  71 022308.
\bibitem{15} Zhou L,  Song H  S,  Guo Y Q, and  Li C, 2003 Phys. Rev. A
 68 024301.
\bibitem{16} Fu H C,Solomon A I, Wang X G, 2002 J. Phys. A 35 4293.
             Li S B, Xu J B, 2005 Phys. Lett. A  334 109.
\bibitem{17} Cheng W W, Huang Y X, Liu T K and Li H, 2007 Physica E 39 150.
\bibitem{18} Xin R, Song Z, Sun C P, 2005 Phys. Lett. A 342 30.
\bibitem{19} Huang Z, Osenda O, Kais S, 2004 Phys. Lett. A 322  137.
\bibitem{20} Osenda O, Huang Z, Kais S, 2003 Phys. Rev. A 67  062321.
\bibitem{21} Osterloh A, Amico L, Falci G, and  Fazio R, 2002 Nature(London) 416 608.
\bibitem{22} Lieb E, Schultz T, Mattis D, 1961 Ann. Phys. 60  407.
\bibitem{23} Wick G C, 1950  Phys. Rev. 80 268.
\bibitem{24} Wooters W K, 1998 Phys. Rev. Lett. 80  2245.
\bibitem{25} Osborne T J, Nielsen M A, 2002 Phys. Rev. A 66  032110.

\end{thebibliography}
\end{document}